\documentclass{article} 
\usepackage[]{graphicx}
\begin{document}

%\preprint{}

\title{Metamaterial-inspired multichannel\\thin-film sensor} 

\date{}
\maketitle 
\begin{center}
Withawat Withayachumnankul$^{1,2}$, Kata Jaruwongrungsee$^{2,3}$,\\Christophe Fumeaux$^{1}$, and Derek Abbott$^{1}$

$^1$School of Electrical \& Electronic Engineering, The University of Adelaide,
Adelaide, SA 5005,
Australia

$^2$School of Electronic Engineering, Faculty of Engineering, King Mongkut's Institute of Technology Ladkrabang, Bangkok 10520, Thailand

$^3$Nanoelectronics and MEMS Laboratory, National Electronics and Computer Technology Center, Pathumthani 12120, Thailand
\end{center}

\begin{abstract} 
A multichannel thin-film sensor is implemented from a set of microstrip-coupled split-ring resonators (SRR’s) with different dimensions. Each SRR exhibits a unique high-Q resonance that is sensitive to the presence of a sample in a particular area. Hence, this SRR-based sensor can function (i) to detect different samples simultaneously to increase the throughput or (ii) to characterise nominally identical samples at multiple frequencies to increase the sensor selectivity. The design principle is validated with simulation and measurement. Owing to the optimized design, sensing a low-permittivity film with a thickness as small as one thousandth of the operating wavelength is achievable.
%A high-sensitivity multichannel thin-film sensor is implemented from a set of microstrip-coupled split-ring resonators (SRR’s) with different dimensions. Upon electromagnetic excitation, each SRR exhibits a unique high-Q magnetic resonance that is sensitive to the presence of a sample at a particular sensing location. Hence, this SRR-based sensor can function (i) to detect different samples simultaneously to increase the throughput or (ii) to characterise nominally identical samples at multiple frequencies to increase the sensor selectivity. The design principle is successfully validated with simulation and measurement. Owing to the optimized design, sensing a low-permittivity film with a thickness as small as one thousandth of the operating wavelength is plausible.
\end{abstract}

Split-ring resonators (SRR\rq{}s) are among many fundamental building blocks for metamaterials that can collectively provide customizable values of the permittivity and/or permeability. A SRR is typically made of one or two concentric subwavelength metallic rings (namely single or double SRR, respectively) with a narrow split in each ring \cite{Pen99}. In response to an electromagnetic excitation, it exhibits a strong magnetic resonance whose frequency is determined by its dimensions, geometry, and constituent materials. On resonance, the ring develops an intense and localised electric field at the narrow split. The sensitivity of the resonance frequency to constituent materials, together with the field localisation, subwavelength ring size, and high-Q resonance, makes SRR\rq{}s ideal for thin-film sensing.

Owing to these properties, SRR's have recently been implemented for biosensors in different configurations. Planar arrays of SRR's were employed for thin-film sensing \cite{Deb07,Dri07,Cha10,Chi09,Oha08,Gor11,Tao11}. However, those sensors require a relatively large amount of the sample to uniformly cover an array with tens to hundreds of SRR\rq{}s during the measurement. The constraints of the sample size and uniformity can be eliminated by using either microstrip-coupled SRR's \cite{Lee08,Lee08b,Lee09} or waveguide-loaded SRR\rq{}s \cite{Aln08}. Although both microstrip and waveguide arrangements can be used for thin-film sensing, electromagnetic coupling is much stronger in the former case, resulting in its higher immunity against noise and fluctuations.

%It was shown that biomolecules covering a few number of double SRR's can be detected by using this microstrip configuration \cite{Lee08,Lee08b,Lee09}. 

For the microstrip-coupling configuration, an SRR is positioned in a close proximity to a microstrip transmission line, which builds up a magnetic field around itself in a quasi-TEM wave propagation. This oscillating magnetic field induces circulating current in the SRR loop. In the quasistatic limit, an SRR can be approximated by an inductor and a capacitor in the form of a series \textit{LC} resonant circuit. Specifically, the ring forms the inductor, and the split forms the capacitor. The resonance takes place in the SRR when the electric energy stored in the capacitor is balanced with the magnetic energy stored in the inductor. Loading a sample onto the SRR surface alters the total capacitance of the structure and hence results in a detectable change in the resonance frequency.

This article presents a multichannel thin-film sensor based on microstrip-coupled SRR's. Several SRR's with different sizes are placed along a microstrip line. Thus, the sensor exhibits multiple resonances, each of which represents a single channel governed by an individual SRR. Each resonance is sensitive to the presence of a sample deposited at the narrow split of the corresponding SRR. Since every resonator possesses a high-Q resonance at a unique frequency, mutual interaction among the resonators is minimal. Therefore, several SRR\rq{}s can be placed densely without compromising the performance. %As important as the multichannel sensing capability, this sensor employs single SRR's that are optimal for detecting a change in the small split region. Furthermore, in this article the linearity of each SRR with respect to the sample volume and dielectric constant is fully characterised for the first time. 

\begin{figure}
\centering
\includegraphics{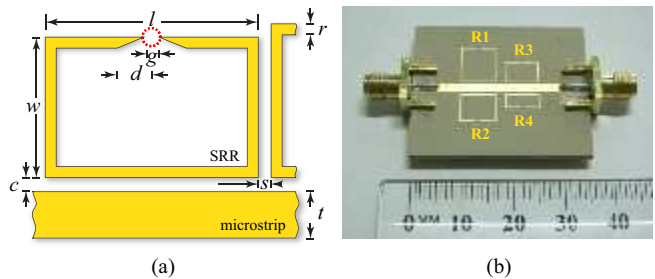}
\caption{(a) Diagram (not to scale) for each SRR, placed alongside a microstrip line, in the multichannel sensor. The sensing area is indicated by the red dotted circle. (b) Fabricated four-channel sensor containing four SRR\rq{}s.}
\label{fig:BA_sensor}
\end{figure}

Beyond the multichannel sensing capability, this article proposes additionally a strategy to improve significantly the sensitivity of SRR sensors by creating a strong and localized field enhancement in a single spot. In earlier designs, several double SRR’s \cite{Lee08,Lee08b,Lee09} or a single SRR with two splits \cite{Aln08} have been used. In such structures, several distributed capacitors collectively determine the resonance frequency, and a considerable amount of sample is required to change all of these capacitors. If the sample partially covers  the structure, the unaffected capacitance reduces the sensitivity in the sensing region as the capacitive change is then not as pronounced. In the present design, each resonance is caused by a single SRR with only one split, as shown in Fig.~\ref{fig:BA_sensor}(a), and consequently, only one distributed capacitor is present. A change in this capacitor by sample loading in the small split region has a direct and large impact on the resonance frequency.

%In additional to the multichannel sensing capability, the sensitivity of the proposed sensor is significantly improved. Here, each resonance is caused by a single SRR with only one split, as shown in Fig.~\ref{fig:BA_sensor}(a). This design is different from several double SRR\rq{}s used in \cite{Lee08,Lee08b,Lee09} or a single SRR with two splits used in \cite{Aln08}. In those earlier designs, there exist several distributed capacitors in the structure that dominate the resonance frequency. Hence, a considerable amount of sample is required to change all of these capacitors. Otherwise, the unaffected capacitors downplay a capacitive change in the sensing area and therefore significantly reduce the sensitivity. For a single SRR with a single split, only one distributed capacitor is present. 

Fig.~\ref{fig:BA_sensor}(a) also depicts some other important features of the implemented resonators. Each SRR has a rectangular shape to maximize coupling to the microstrip transmission line. This leads to a stronger resonance that is robust to measurement uncertainties. At each SRR split, sharp tips are adopted to concentrate the electric field to a small spot \cite{Aln08}. In terms of a series \textit{LC} circuit, the tapered shape of the tips decreases the capacitance at the split, and hence increases the resonance Q-factor. 

Fig.~\ref{fig:BA_sensor}(b) shows the proposed sensor fabricated by using standard photolithography and chemical etching. This particular four-channel sensor is realized from four different SRR\rq{}s positioned along a 50-$\Omega$ microstrip line. All of the resonators share the same dimensions excepting the width, $w$, that equals 8, 6, 4.5, and 3~mm for R1, R2, R3, and R4, respectively. This causes a difference in the loop inductance and hence in the resonance frequency among the SRR\rq{}s. The other dimensions common to all of the SRR\rq{}s are as follows: $l=7$~mm, $g=0.15$~mm, $d=0.525$~mm, $r=0.2$~mm, $c=0.65$~mm, $s=2$~mm, and $t=1.7$~mm. The metal for the SRR\rq{}s, microstrip, and ground plane is copper with a thickness of 35~$\mu$m, coated with gold to prevent oxidization. The substrate is an RT/duroid 6010.2LM high-frequency laminate (ceramic-PTFE composite) with a thickness of 1.90~mm, a relative permittivity of 10.2, and a loss tangent of 0.0023. This sensor specification yields four high-Q resonances in the microwave S band.

\begin{figure}
\centering
\includegraphics{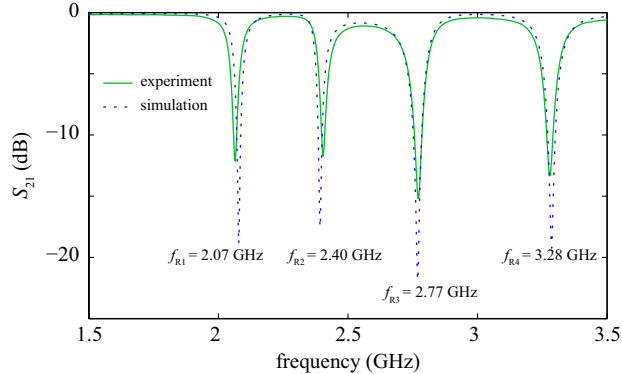}
\caption{Transmission profiles from simulation and experiment with no sample loading. Each resonance, i.e., channel, corresponds to an individual SRR. The indicated resonance frequencies are extracted from the experimental results.}
\label{fig:BA_s21ref}
\end{figure}

During the experiment, the sensor is connected to a vector network analyzer, Agilent Technologies N5230A, and the transmission parameter $S_{21}$ is registered at room temperature in the range between 1.5--3.5~GHz with a resolution of 3~MHz. The numerical simulation is carried out with a commercial full-wave electromagnetic solver, CST Microwave Studio, based on the Finite-Integration Technique (FIT). The gap size $g$ in each ring resonator is slightly adjusted in the simulation to reflect the fabrication imperfections, as observed under a microscope.

The measured and simulated transmission profiles of the sensor without any sample are shown in Fig.~\ref{fig:BA_s21ref} with small discrepancies caused by fabrication tolerances and limits in simulation accuracy. The results show four non-overlapping resonances corresponding to the four distinct SRR\rq{}s. The lowest resonance $f_{\rm R1}$ is associated with R1, which has the largest inductance, the second lowest resonance $f_{\rm R2}$ with R2, and so on. No change in the resonances is observed during 5 hours of measurement, demonstrating the thermal stability of the sensor.

\begin{figure}
\centering
\includegraphics{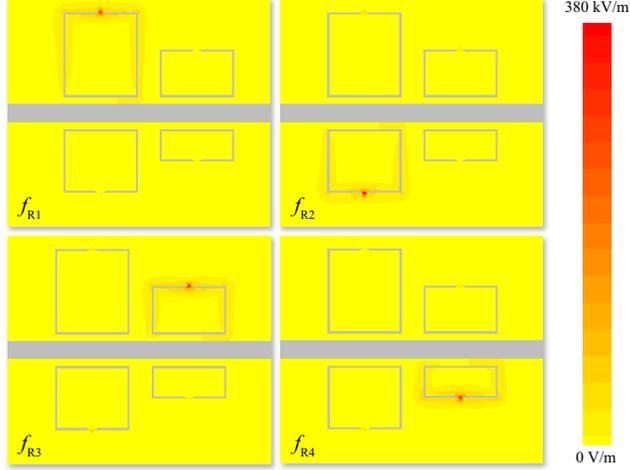}
\caption{Simulated electrical-field amplitude distribution on the sensor surface at four resonance frequencies. The peak input power supplied to the sensor is 1~W.}
\label{fig:BA_field}
\end{figure}

Further insight can be obtained from the simulation, as shown in Fig.~\ref{fig:BA_field}, where the electric-field distribution on the sensor surface is calculated for every channel. At each resonance frequency, only the relevant SRR establishes a strong and highly confined electric field around the capacitive split, and no field enhancement is observed elsewhere. The field enhancement in any active ring is as high as 380~kV/m, given the input power of only 1~W. These calculated field distributions confirm the extreme field confinement achieved with the implemented SRR\rq{}s, which is a basis for high-performance sensing.

\begin{figure}
\centering
\includegraphics{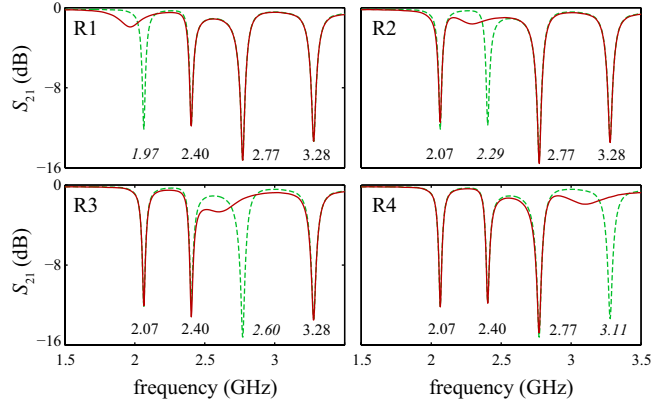}
\caption{Transmission profiles for unloaded (dotted lines) and loaded (solid lines) SRR\rq{}s. Each subfigure represents a comparison of sample loading at an individual SRR channel. The specified numbers are the resonance frequencies in GHz of the loaded sensor, and the italization indicates shifted resonances.}
\label{fig:BA_s21mult}
\end{figure}

In the next experiment, ethanol is used as a sample under test by the multichannel sensor. At 3~GHz, ethanol has a relative permittivity and loss tangent of 6.0 and 0.96, respectively \cite{Mud74}. In each measurement, 5 $\mu$L of ethanol is dropped by using a high-precision pipette onto the split of a selected SRR channel, and the sensor\rq{}s response is observed. The transmission results are shown in Fig.~\ref{fig:BA_s21mult}. The observed resonance shifting and damping indicate the presence of the sample in the selected channel. By comparing the resonance frequencies of the loaded sensor in Fig.~\ref{fig:BA_s21mult} with the reference values in Fig.~\ref{fig:BA_s21ref}, it is clear that non-active SRR channels show no detectable change in the resonance frequency. Hence, it is clear that the four sensing channels can operate independently. 

Each SRR can be approximated by a series \textit{LC} circuit with a resonance frequency $f_0=1/(2\pi\sqrt{LC})$, where $L$ denotes the ring inductance and $C$ denotes the split capacitance. The sensitivity of the resonance frequency to the split capacitance can be expressed as \cite{Wit11}
\begin{eqnarray}
S_{C}^{f_0} = \frac{C}{f_0}\frac{\partial f_0}{\partial C} = -0.5\;.
\end{eqnarray}
It can be deduced from this equation that if the relative change in the capacitance formed by the split, $\Delta C/C$, is constant, the relative change in the resonance frequency, $\Delta f/f_0$, is fixed, regardless of the ring inductance. Nonetheless, the relative change in the resonance frequency for the active channel varies between -4.6\% and -6.1\% among the four channels, despite a fixed volume of sample. This small variation is presumably caused by manual sample deposition, which results in a slight difference in the sample location. This uncertainty can be readily overcome with an automated process.

The sample volume used in the experiment is considerably small. Nevertheless, additional simulations suggest that only one nanolitre of ethanol, i.e., a 0.1-mm cube, is unambiguously detectable with a frequency shift of 15~MHz. In other words, the ratio between the operating wavelength and the film thickness can be as large as three orders of magnitude. %In addition, ethanol is used in the experiment instead of biological substances for the sake of reproducibility. The sensor relies on a change in the dielectric constant, which is equivalent for either pure liquid or a biological sample. In this case, the obtainable results well demonstrate the sensing and multichannel capability. Experiments with biological materials can be performed with the implemented sensor, but this does not necessarily reflect its performance.

In conclusion, this article presents a thin-film sensor with a series of optimized SRR's as sensing elements. The fabricated sensor successfully performs multichannel detection with an enhanced sensitivity. Scaling down the structure to operate at higher frequencies further reduces the minimum amount of detectable sample. More channels can be added to the sensor provided that resulting resonances are sparse enough to avoid mutual coupling. The resonance of each channel can be positioned at an abritary frequency where unique dielectric features of the sample are expected. The realization can be used for on-site disposable sensors, which allow sensing with either high selectivity or high throughput.

\section*{Acknowledgments}

The authors acknowledge Pavel Simcik and Henry Ho for their technical assistance. The laminates used in the experiment were supplied by Rogers Corporation. This research was supported by the Australian Research Council \textit{Discovery Projects} funding scheme (project number DP1095151).

%\bibliography{2011_BA_APL}

\begin{thebibliography}{10}

\bibitem{Pen99}
J.~B. Pendry, A.~J. Holden, D.~J. Robbins, and W.~J. Stewart, ``Magnetism from
  conductors and enhanced nonlinear phenomena,'' {\em IEEE Trans. Microwave
  Theory Tech.}~{\bf 47}(11), pp.~2075--2084, 1999.

\bibitem{Deb07}
C.~Debus and P.~H. Bolivar, ``Frequency selective surfaces for high sensitivity
  terahertz sensing,'' {\em Appl. Phys. Lett.}~{\bf 91}, p.~184102, 2007.

\bibitem{Dri07}
T.~Driscoll, G.~O. Andreev, D.~N. Basov, S.~Palit, S.~Y. Cho, N.~M. Jokerst,
  and D.~R. Smith, ``Tuned permeability in terahertz split-ring resonators for
  devices and sensors,'' {\em Appl. Phys. Lett.}~{\bf 91}, p.~062511, 2007.

\bibitem{Cha10}
Y.-T. Chang, Y.-C. Lai, C.-T. Li, C.-K. Chen, and T.-J. Yen, ``A
  multi-functional plasmonic biosensor,'' {\em Opt. Express}~{\bf 18}(9),
  pp.~9561--9569, 2010.

\bibitem{Chi09}
S.-Y. Chiam, R.~Singh, J.~Gu, J.~Han, W.~Zhang, and A.~A. Bettiol, ``Increased
  frequency shifts in high aspect ratio terahertz split ring resonators,'' {\em
  Appl. Phys. Lett.}~{\bf 94}(6), p.~064102, 2009.

\bibitem{Oha08}
J.~F. O'Hara, R.~Singh, I.~Brener, E.~Smirnova, J.~Han, A.~J. Taylor, and
  W.~Zhang, ``Thin-film sensing with planar terahertz metamaterials:
  sensitivity and limitations,'' {\em Opt. Express}~{\bf 16}(3),
  pp.~1786--1795, 2008.

\bibitem{Gor11}
J.~A. Gordon, C.~L. Holloway, J.~Booth, S.~Kim, Y.~Wang, J.~Baker-Jarvis, and
  D.~R. Novotny, ``Fluid interactions with metafilms/metasurfaces for tuning,
  sensing, and microwave-assisted chemical processes,'' {\em Phys. Rev. B:
  Condens. Matter}~{\bf 83}, p.~205130, 2011.

\bibitem{Tao11}
H.~Tao, L.~Chieffo, M.~A. Brenckle, S.~M. Siebert, M.~Liu, A.~C. Strikwerda,
  K.~Fan, D.~L. Kaplan, X.~Zhang, R.~D. Averitt, and F.~G. Omenetto,
  ``Metamaterials on paper as a sensing platform,'' {\em Adv. Mater.} , 2011.
\newblock (Online edition).

\bibitem{Lee08}
H.-J. Lee and J.-G. Yook, ``Biosensing using split-ring resonators at microwave
  regime,'' {\em Appl. Phys. Lett.}~{\bf 92}, p.~254103, 2008.

\bibitem{Lee08b}
H.-J. Lee, H.-S. Lee, K.-H. Yoo, and J.-G. Yook, ``On the possiblity of
  biosensors based on split ring resonators,'' in {\em Proceedings of the 38th
  European Microwave Conference},  pp.~1222--1225, 2008.

\bibitem{Lee09}
H.-J. Lee, H.-S. Lee, K.-H. Yoo, and J.-G. Yook, ``{DNA} sensing based on
  single element planar double split-ring resonator,'' in {\em IEEE MTT-S
  International Microwave Symposium Digest},  pp.~1685--1688, 2009.

\bibitem{Aln08}
I.~A.~I. Al-Naib, C.~Jansen, and M.~Koch, ``Thin-film sensing with planar
  asymmetric metamaterial resonators,'' {\em Appl. Phys. Lett.}~{\bf 93},
  p.~083507, 2008.

\bibitem{Mud74}
R.~E. Mudgett, D.~I.~C. Wang, and S.~A. Goldblith, ``Prediction of dielectric
  properties in oil-water and alcohol-water mixtures at 3000 {MHz}, 25$^\circ$c
  based on pure component properties,'' {\em J. Food Sci.}~{\bf 39},
  pp.~632--635, 1974.

\bibitem{Wit11}
W.~Withayachumnankul, C.~Fumeaux, and D.~Abbott, ``Planar array of
  electric-{LC} resonators with broadband tunability,'' {\em IEEE Antennas
  Wirel. Propag. Lett.}~{\bf 10}, 2011.
\newblock (In press).

\end{thebibliography}
%\bibliographystyle{spiebib}

\end{document}